\documentclass[preprint2]{aastex6}
\usepackage{amssymb,amsmath}
\usepackage{natbib}

\newcommand{\vect}[1]{\mathbf{#1}}
\newcommand{\del}{\nabla}
\newcommand{\unit}[1]{\ensuremath{\, \mathrm{#1}}}
\newcommand{\bigo}{\ensuremath{\mathcal{O}}}
\providecommand{\abs}[1]{\lvert#1\rvert}
\providecommand{\norm}[1]{\lVert#1\rVert}
\newcommand{\figref}[1]{Figure \ref{#1}}

\begin{document}
\title{On excited frequencies for Alfv\'en waves in a coronal arcade}

\author{Lucas A. Tarr\altaffilmark{1,2}}
\affil{National Research Council Research Associate}
\affil{Naval Research Laboratory \\
4555 Overlook Ave SW \\
Washington, DC 20375, USA
}
\altaffiltext{2}{lucas.tarr.ctr@nrl.navy.mil}

\begin{abstract}
  The normal modes of oscillation for a magnetic arcade are used to analytically solve an initial value problem and estimate the power spectra of wave frequencies generated by a reconnection event in the solar corona.  Over a realistic range of parameters, I find that such a disturbance generates a peak power at $\sim10$s$\mathrm{mHz}$ frequencies, but still substantial power up to $\sim4\mathrm{Hz}$.  The cadence and sensitivity of current instrumentation does not allow observations of oscillations at these frequencies, but in the near future, new instrumentation will be able to probe this regime, and observationally determine its energetic importance.
\end{abstract}

\section{Introduction}
Magnetohydrodynamic (MHD) waves likely play an important role in coronal dynamics and heating \citep{vanBallegooijen:2011, Arber:2016,Kerr:2016, Reep:2016}.  Their presence in the corona has been inferred from numerous observations \citep[see the recent review by][and references therein]{Liu:2011}.  Some previous work has determined the normal modes of typical coronal magnetic configurations \citep{Oliver:1993}, but because wave damping is a frequency dependent process \citep{DePontieu:2001, vanBallegooijen:2011} it is also important to determine the frequency spectrum that will be excited by an event.  In this letter I apply the results of an existing theory to analytically estimate the frequency spectrum of Alfv\'en waves in a coronal arcade generated by a local perturbation, such as a small reconnection event.  While the power peaks at low frequency, I do find that a substantial portion of the power generated by such a perturbation will be in frequencies in the $0.5\unit{Hz}<\nu<4\unit{Hz}$ range for a reasonable span of coronal parameters.  Therefore, high cadence observations should be taken to see if such waves are present, and simulations should consider the propagation and damping of waves with these frequencies to determine their importance for basal coronal heating or (possibly) energy flux during flares.

I begin in \S\ref{sec:nm} by rederiving the initial condition for a coronal arcade and the normal modes for perturbations about this equilibrium.  In \S\ref{sec:alfven} I specialize to Alfv\'en waves and determine the power spectrum arising from a small perturbation.  I conclude with some more general considerations in \S\ref{sec:discussion}.

\section{Wave Equation for a Coronal Arcade}\label{sec:nm}
This letter presents a simple application of \citet{Oliver:1993}.  Recent observations of coronal waves and the possibility of high cadence data, particularly from instrumentation at the Daniel K. Inoye Solar Telescope \citet{Tritschler:2016}, make this a worthwhile exercise.  In order to introduce notation and keep this work self contained, I reproduce a condensed portion of \citet{Oliver:1993}, but do refer frequently to that work, throughout.

To begin, I find a magnetohydrostatic equilibrium for a coronal arcade, where the Lorentz, pressure, and gravitational forces are balanced.  Using standard notation, the (static) momentum equation is

\begin{equation}
  \label{eq:momentum}\vect{j}\times\vect{B} - \del P + \rho\vect{g} = \vect{0}.
\end{equation}

After taking the scalar product of \eqref{eq:momentum} with $\vect{B}$, I get

\begin{equation}
  \label{eq:p1}-B\frac{d P}{d s} - \rho B_z g = 0,
\end{equation}

where $s$ is a parameter along the field.  I assume the coronal plasma is isothermal, fully ionized, and satisfies the gas ideal law

\begin{equation}
  \label{eq:ideal}P = \rho k_B T/\mu
\end{equation}

with $\mu$ the mean atomic mass.  Eqs.~\eqref{eq:p1} and \eqref{eq:ideal} imply

\begin{equation}
  \frac{dP}{dz} = -P\frac{\mu g}{k_BT}
\end{equation}

so that

\begin{equation}
  P(z) = P_0 e^{-z/\Lambda} \text{ and } \rho(z) = \rho_0 e^{-z/\Lambda},
\end{equation}

where $\Lambda = \frac{k_BT}{\mu g}$ is the pressure scale height.  For a low $\beta (=2\mu_0P/B^2)$ plasma and fluid displacements $\pmb{\xi}<<\Lambda$, the momentum equation for static equilibrium gives $\vect{j}\times\vect{B} = \vect{0}$.  One solution is a potential field with $\del\times\vect{B}=\vect{0}$.  Together with the solenoidal condition, and taking the $\hat{\vect{y}}$ direction to be invariant, I can write the potential field as

\begin{gather}
  \vect{B} = \del A(x,y)\times\vect{\hat{y}} = \Bigl(-\frac{\partial A}{\partial z},0,\frac{\partial A}{\partial x}\Bigr).
\end{gather}

The flux function $A$ satisfies Laplace's equation $\del^2 A = 0$.  I solve it via separation of variables with the boundary conditions that $A(x=0)= 0$ and $A(z\rightarrow\infty) = 0$.  The solution is

\begin{gather}
  A(x,y) = B_0\Lambda_B\cos\bigl(\frac{x}{\Lambda_B}\bigr)e^{-\frac{z}{\Lambda_B}}
  \intertext{from which}
  B_x = B_0\cos\Bigl(\frac{x}{\Lambda_B}\Bigr)e^{-\frac{z}{\Lambda_B}}\\
  B_z = -B_0\sin\Bigr(\frac{x}{\Lambda_B}\Bigl)e^{-\frac{z}{\Lambda_B}}
\end{gather}

$\Lambda_B$ is the magnetic scale height and is related to the width of the arcade through $\Lambda_B = \frac{2L}{\pi}.$  

Define $\delta $ as the ratio of magnetic to pressure scale heights,

\begin{equation}
  \delta = \frac{\Lambda_B}{\Lambda}.
\end{equation}

Ignoring gravity amounts to setting $\Lambda=\infty$ so that $\delta = 0$; on the other hand, $\delta = 2$ when the magnetic pressure and plasma pressure have the same scale height.  The latter scenario also sets $\beta = $constant.  \citet{Oliver:1993} point out that $\delta\propto L/T$, so for coronal values of a $100\unit{Mm}$ loop at $1\unit{MK}$, the ratio is $\delta\approx 1.05$.

My goal is to determine the response of the above equilibrium to an initial perturbation.  I find the arcade's normal modes in terms of the displacement field, $\pmb{\xi}$.  Note the modes determined below are the same as \citet{Oliver:1993}'s velocity modes, with $\vect{v}=\partial_t\pmb{\xi}$.  

The wave equation is derived in a standard way \citep{Priest:1982}: the perturbations are taken to be adiabatic, which removes the energy equation from consideration, and  the induction and continuity equations are substituted in the time derivative of the momentum equation.  To make the usual coronal approximation I assume that, for the perturbations, the pressure and gravitational forces are negligible compared to magnetic forces.  Finally, I drop all terms $\bigo(2)$ and arrive at the linearized wave equation:

\begin{equation}
  \label{eq:wave1}\rho_0\frac{\partial^2\pmb{\xi}}{\partial t^2} = \frac{1}{\mu_0}\{\del\times[\del\times(\pmb{\xi}\times\vect{B}_0)]\}\times\vect{B}_0.
\end{equation}

The background state is curl--free and invariant in the $\vect{\hat{y}}$ direction, with $\vect{B}_0=\del A\times\vect{\hat{y}}$.  An appropriate coordinate system has unit vectors

\begin{equation}
  \vect{b} = \vect{e}_\parallel = \frac{\vect{B}_0}{\abs{\vect{B}_0}},\ \vect{e}_\perp = \vect{e}_y,\ \vect{e}_n = \frac{\del A}{\norm{\del A}}.
\end{equation}

With a bit of algebra, the wave equation \eqref{eq:wave1} may be expressed in that coordinate system.  Lastly, I assume harmonic time dependence for each mode, so $\pmb{\xi}(\vect{x},t)\rightarrow\pmb{\xi}(\vect{x})e^{-i\omega t}$.  The result is

\begin{equation}
  \label{eq:wave2}-\rho\omega^2\pmb{\xi} = \frac{1}{\mu_0}(\vect{B}_0\cdot\del)^2\xi_y\vect{\hat{y}}+\frac{1}{\mu_0}[\del^2(\pmb{\xi}\cdot\del A)]\del A
\end{equation}

where, for a general scalar function $f$,

\begin{equation}
  (\vect{B}_0\cdot\del)^2f = [(\vect{B}_0\cdot\del)](\vect{B}_0\cdot\del f).
\end{equation}

\section{Solution for Alfv\'en Waves}\label{sec:alfven}
The wave equation \eqref{eq:wave2} has solutions corresponding to generalized Alfv\'en and fast waves (the slow waves were removed by ignoring pressure forces).  Assuming the perturbations themselves are invariant in $\vect{e}_y$ (a rather severe restriction; see the Discussion) decouples the Alfv\'en and fast modes.  I focus on the Alfv\'en waves to keep the present work analytic.  The Alfv\'en waves are polarized in the perpendicular direction $\vect{e}_y$, and therefore only involve the component $\xi_y$ (when needed, I notate the in--plane field $\xi_b$):

\begin{equation}
  \label{eq:wavealfven}-\rho\omega^2\xi_y = \frac{1}{\mu}(\vect{B}_0\cdot\del)^2\xi_y.
\end{equation}

The gradient along the field is found using the fact that the potential is constant along field lines.  For a field line identified by $\vect{x} = (x_0,0)$,

\begin{align}
  A(x,z) & = B_0\Lambda_B\cos\frac{x}{\Lambda_B}\exp(-\frac{z}{\Lambda_B}) \\
  & = A(x_0,0) = B_0\Lambda_B\cos\frac{x_0}{\Lambda_B}.
\end{align}

Along that field line, $z$ are $x$ are related by

\begin{gather}
  \label{eq:xz}\cos\frac{x}{\Lambda_B} = \cos\frac{x_0}{\Lambda_B}\exp(\frac{z}{\Lambda_B}),
  \intertext{from which}
  \label{eq:dzdx}\frac{\partial z}{\partial x} = -\frac{\sin\frac{x}{\Lambda_B}}{\cos\frac{x_0}{\Lambda_B}}\exp \frac{z}{\Lambda_B} = \frac{B_z}{B_0\cos\frac{x_0}{\Lambda_B}}.
\end{gather}

Equation \eqref{eq:xz} is of the form $f=f(x,z(x))$.  Taking the total derivative in $x$ of \eqref{eq:xz}, and using \eqref{eq:dzdx} and the definition of $B_x$, I write the derivative along the field as

\begin{equation}
  \vect{B}_0\cdot\del = B_x\frac{d}{dx} = \frac{A(x_0,0)}{\Lambda_B}\frac{d}{dx}.
\end{equation}

Substituting the above into \eqref{eq:wavealfven} and rearranging terms, I finally arrive at an ordinary differential equation for the displacement field $\xi_y(x)$,

\begin{gather}
  \frac{d^2\xi_y}{dx^2}+\frac{\omega^2\rho_0}{B_0^2/\mu_0}e^{-\delta\frac{x}{\Lambda_B}}\cos^{-2}\frac{x_0}{\Lambda_B}\xi_y = 0,
\intertext{which is more usefully written}
\label{eq:wave3}\frac{d^2\xi_y}{dx^2}+\frac{\omega^2}{V_{A0}}\Biggl[\frac{\cos\frac{x_0}{\Lambda_B}}{\cos\frac{x}{\Lambda_B}}\Biggr]^\delta\cos^{-2}\Bigl(\frac{x_0}{\Lambda_B}\Bigl)\xi_y = 0
\end{gather}

with $V_{A0}^2 = V_{A}(z=0)^2$.

An analytic solution exists when the pressure scale height is much larger than the magnetic scale height, so $\delta\rightarrow 0$.  This is the case when gravity is ignored.  Then \eqref{eq:wave3} has constant coefficients and describes simple harmonic oscillation,

\begin{equation}
  \frac{d^2\xi_y}{dx^2}  = -k_x^2\xi_y,
\end{equation}

with solutions

\begin{gather}
  \xi_y(x) = \begin{cases}
    \frac{1}{x_0}\cos k_xx & \text{even function}\\
    \frac{1}{x_0}\sin k_xx & \text{odd function}
  \end{cases}\\
\text{where }  k_x^2 = \frac{\omega^2}{V_{A0}^2}\cos^{-2}\frac{x_0}{\Lambda_B}.
\end{gather}

Let the field be line--tied at the lower boundary (e.g. at $x=\pm x_0$) so $\xi_y(x=\pm x_0)=0$.  The allowed normal modes are discrete:

\begin{align}
  k_x^{(n)} = \frac{(n+\frac{1}{2})\pi}{x_0},\  & n = 0,1,2,\ldots \text{ (even) }\\
  k_x^{(m)} = \frac{m\pi}{x_0},\  & m = 1,2,\ldots \text{ (odd) },
\end{align}

each with corresponding eigenfunction $\xi_y^{(n)}(x)$ or $\xi_y^{(m)}(x)$.  Even and odd refer to the parity of the perturbation.  Note that for a given field line the oscillatory modes are discrete, but this discrete spectrum shifts continuously from field line to field line.  \citet{Oliver:1993} stress that, taken as a whole, the arcade can be considered a system with a \emph{continuous} frequency spectrum.  I would rather stress that the result of an observation may depend on orientation.  A perfect observer analyzing an optically thin plasma would find discrete normal modes when aligned with the arcade axis (perhaps near the solar limb), but a continuous spectrum when viewed from above (near disk center), provided she could somehow detect transverse Alfv\'enic oscillations in that case.  Either way, there is a case to be made for studying the center--to--limb variation of oscillatory power in coronal arcades.

\begin{figure}
  \includegraphics[width=0.5\textwidth]{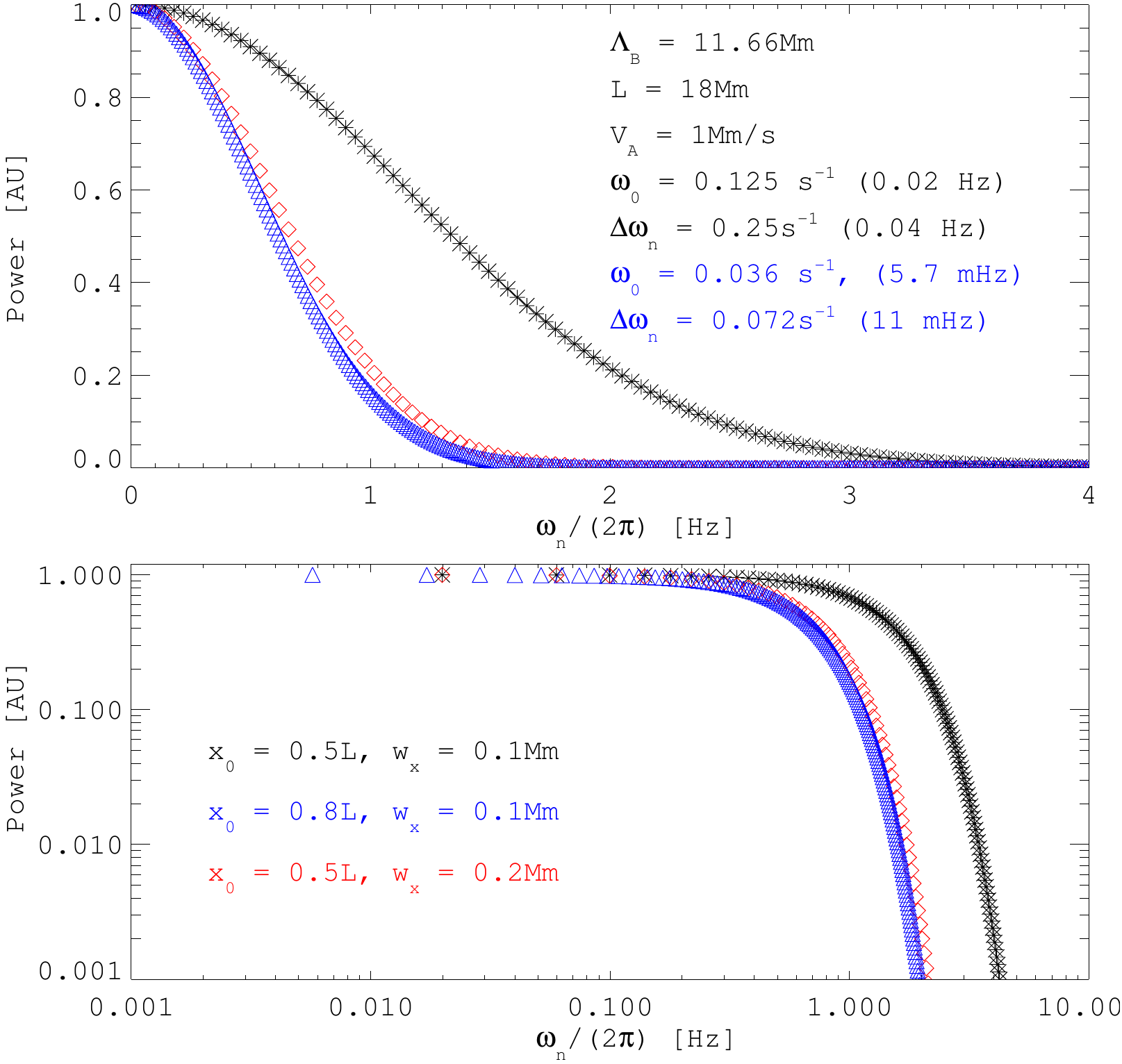}\caption{\label{fig:alfven-d0}Power as a function of frequency for a Gaussian initial condition.  Top: linear scaling.  Bottom: log scaling.}
\end{figure}

To determine which modes are excited for an initial perturbation, I impose a Gaussian perturbation at the top of a loop, characterized by some width $w_x$:

\begin{equation}
  \xi_y(x)=\frac{1}{\sqrt{2\pi}w_x}e^{-\frac{x^2}{2w_x^2}}.
\end{equation}

This is an even function, so I project it onto the even eigenfunctions of the system:

\begin{equation}
  \xi_y(x) = \sum_{n=0}^{\infty}A_n\xi_y^{(n)}(x).
\end{equation}

  The coefficients $A_n$ are determined in the standard way,

\begin{equation}
  A_n = \int\xi_y(x)\xi_y^{(n)}(x)dx.
\end{equation}

The integral of a Gaussian and $\cos(x)$ has normalized solution \citep[][\S7.4.6]{Abramowitz:1964}

\begin{equation}
  \label{eq:coeff}A_n = \frac{1}{x_0}e^{-\frac{1}{2}\frac{(n+\frac{1}{2})^2\pi^2w_x^2}{x_0^2}} = \frac{1}{x_0}e^{-\frac{1}{2}\frac{w_x^2\omega_n^2}{V_{A0}^2\cos^2(x_0/\Lambda_B)}}
\end{equation}

The power in each frequency is given by $\abs{A_n}^2$, and Equation \eqref{eq:coeff} shows how the power changes as the arcade and perturbation properties are varied.  The full solution is

\begin{equation}
\xi_y(x,t) = \sum_{n=0}^\infty \frac{1}{x_0}\exp\Bigl[-\frac{(n+\frac{1}{2})^2\pi^2w_x^2}{2x_0^2}\Bigr]\cos\frac{(n+\frac{1}{2})\pi x}{x_0}
\end{equation}

\figref{fig:alfven-d0} shows the first several hundred coefficients for the power ($\abs{A_n}^2$) for three solutions: $(x_0/L, w_x/\unit{Mm}) = (0.5,0.1)$ (black), $(0.8,0.1)$ (blue), and $(0.5,0.2)$ (red).  To fix the physical parameters I used the flare observations reported by \citet{Jing:2016} to estimate the width of an arcade as the distance between flare ribbons, $L\approx 18\unit{Mm}$, and the width of the perturbation as the leading edge of flare ribbons, $w_x\approx 0.15\unit{Mm}$.  Recall that $L$ is the half--width in the present model.  An arcade full--width of $36\unit{Mm}$ also roughly agrees with the well observed post--flare arcade from the Bastille Day flare in 2000 \citep{Somov:2002}.  \citet{Morton:2013} analyzed EUV data from the Hi--C sounding rocket and found that loop structures supporting Alfv\'en waves have a cross section of $\approx150\unit{km}$, which supports using a perturbation width of that size.

Comparing the black and blue series in \figref{fig:alfven-d0} shows the effect of holding the size of the perturbation fixed and changing the length of the loop on which it is introduced.  The larger loop (blue: greater $x_0$, further from the arcade axis) has a closer spacing between the excited frequencies in a give range, but a lower fundamental frequency.  For these parameters, the outer portions of the arcade $(x_0\gtrsim0.8 L)$ have fundamental frequencies that approach the p--mode spectrum, allowing for another possible excitation mechanism that I do not explore further here.

Comparing the red and black series shows the effect of changing the size of the perturbation.  For the same size loop, the larger length perturbation (red) concentrates the power at lower frequencies.  Observationally determining the oscillatory spectrum of loops with a known size may help constrain the size of the driver, for instance the extent of a current sheet in the Parker braiding model.  On the other hand, if a perturbation were to maintain a given size and excite loops in an arcade at progressively greater heights then one would expect to see a drift towards lower frequencies in observations of oscillatory power.  For the several cases considered here, the generated spectrum for a perturbation using realistic coronal values contains substantial power up to $\approx 1\unit{Hz}$, and can contain power up to $3\unit{Hz}$ or so.

\section{Discussion}\label{sec:discussion}
In the present work I have focused on analytic results for Alfv\'en waves in a coronal arcade, which are only available when $\delta=0$.  In the introduction I stated that under normal coronal conditions we expect $\delta\approx1$.  Figure $2b$ of \citet{Oliver:1993}, which I have also verified, shows that there is little difference between the fundamental frequency for a given loop of the arcade between $0<\delta<2$: the frequencies shift slightly upward with increasing $\delta$, but that is the only major difference; higher harmonics follow the same trend.

I have also ignored waves excited in the plane of the magnetic field, which are fast mode waves.  The distribution of excited frequencies for the fast mode is a more involved problem than for Alfv\'en waves and outside the scope of this short note; however the total power excited in Alfv\'en compared to compressible modes is easier to estimate because, owing to the translational invariance in $y$, the arcade system decouples the in--plane and out--of--plane directions.  The power excited in Alfv\'en waves is $\sim \abs{\xi_y}^2$, the power in fast waves is $\sim\abs{\xi_b}^2$, and for an ensemble of randomly oriented perturbations each type of wave would receive half the power.  Their behavior is markedly different, though.  The Alfv\'en waves, constrained to given sets of field lines, maintain a concentrated power, while fast mode waves spread that power out as they refract, as demonstrated in the simulations in \citet{Russell:2013}.

A greater restriction is the translational invariance of the perturbation itself, which is unlikely to result from reconnection.  Relaxing that assumption would recouple the Alfv\'en and fast modes, and it is unclear how the energy will ultimately partition into each mode.  The coupled problem is analytically tractable in certain situations, for instance by assuming a density profile that traps the fast waves \citep{Hindman:2015}.  3D MHD simulations may also be used to tackle the problem \citep{Rial:2010}, but it should be noted that high frequency phenomena come hand--in--hand with high wavenumber and thus small spatial scales.  High resolution simulations are expensive to perform, and any power that would be generated at higher frequencies is unresolved: the authors just mentioned were able to detect power in just the first several modes in a 3D numerical experiment of this same arcade system.  These difficulties in turn suggested studying this simple, analytically tractable case in the first place.  For now, I simply accept the limitations and take the present results as evidence that high frequency waves should exist, and as a first--pass estimation of their spectrum, in coronal arcades.

There is already some observational evidence of high frequency waves on the Sun.  \citet{Deforest:2004} analyzed TRACE 1600\AA{} data and found frequencies  $\nu\approx100\unit{mHz}$.  Although emission in the 1600\AA{} channel forms in the chromosphere, the presence of high frequency waves in the TRACE data, combined with lower reflection coefficients for higher frequency waves, supports testing for their presence in the low corona.  Looking to the future, both the DLNIRSP and CryoNIRSP instruments at the upcoming DKIST observatory should have the sensitivity and cadence to probe the high frequency component of the coronal wave field and determine what its energetic import may be.  Given the results of this most--simple model, the observations may be able to detect shifts in oscillatory power dependent on position within an arcade, or through center--to--limb variations, that will provide insight into the presence of high frequency coronal waves and possible excitation mechanisms.

\acknowledgements  This work was supported by the Chief of Naval Research, and carried out while LAT was a National Reseach Council Research Associate at the United States Naval Research Laboratory.

\end{document}